\documentclass[12pt]{article}
\begin{document}
\begin{titlepage}
\begin{flushright}
hep-th/0404250\\
TIT/HEP-522\\
TU-718\\
April, 2004\\
\end{flushright}
\vspace{0.5cm}
\begin{center}
{\Large \bf Singlet Deformation and Non(anti)commutative \\
${\cal N}=2$ Supersymmetric $U(1)$ Gauge Theory
}
\lineskip .75em
\vskip2.5cm
{\large Takeo Araki}${}^{1}$\ \
and \ \ 
{\large Katsushi Ito}${}^{2}$ \
\vskip 2.5em
${}^{1}$ {\large\it Department of Physics\\
Tohoku University\\
Sendai, 980-8578, Japan}  \vskip 1.5em
${}^{2}${\large\it Department of Physics\\
Tokyo Institute of Technology\\
Tokyo, 152-8551, Japan}  \vskip 4.5em
\end{center}
\begin{abstract}
We study ${\cal N}=2$ supersymmetric $U(1)$ gauge theory
in non(anti)commutative ${\cal N}=2$ harmonic superspace
with the singlet deformation, which preserves chirality.  
We construct a Lagrangian which is invariant under both the deformed 
gauge and supersymmetry transformations.
We find the field redefinition such that the ${\cal N}=2$ vector
 multilplet transforms canonically under the deformed symmetries.

\end{abstract}
\end{titlepage}
\baselineskip=0.7cm
Non(anti)commutative superspace\cite{ncsuper} with
nonanticommutativity in Grassmann odd coordinates appears
in superstrings compactified on Calabi-Yau threefold 
in the graviphoton background \cite{OoVa,BeSe,DeGrNi}.
The low-energy effective theory on the D-brane is realized 
by supersymmetric gauge theories in non(anti)commutative superspace.
Perturabative and non-perturabative aspects of these gauge theories have
been studied extensively\cite{Se,ArItOh1,Pert,Inst}.
It is an interesting problem to study the deformation of extended
superspace since it admits a variety of deformation parameters\cite{KlPeTa}.
The deformation of extended superspace has been
recently studied in \cite{IvLeZu,FeSo,ArItOh2,SaWo,Mi,KeSa}. 
In a previous paper \cite{ArItOh2}, we have studied the deformed Lagrangian
explicitly in the component formalism up to the first order in the
deformation parameter $C$. 
Since the Lagrangian get higher order correction in $C$, the full 
Lagrangian is rather complicated.
Moreover, the harmonic superspace formalism introduces the infinite number
of auxiliary fields. 
In order to preserve the WZ gauge in the deformed theory, the gauge
transformation has also correction in the form of 
power series in $C$.

There exist some interesting cases where the deformation structure
becomes simple.
One is the limit to the ${\cal N}=1/2$ superspace\cite{Se}, where the 
action should reduce to ${\cal N}=1/2$ super Yang-Mills theory with 
adjoint matter.
Another interesting case is the singlet deformation\cite{IvLeZu,FeSo}, 
where the deformation parameters
belongs to the singlet representation of the $R$-symmetry group $SU(2)_R$.
In this paper, we will study
${\cal N}=2$ supersymmetric $U(1)$ gauge theory in  the 
harmonic superspace with singlet deformation.
In this case, the gauge and supersymmetry
transformations get correction linear in the deformation parameter.
Therefore we can easily perform the field redefinition such that the
component fields transform canonically under the gauge transformation.
In the case of ${\cal N}=1/2$ super Yang-Mills theory, such field
redefinition is also possible\cite{Se}. 
But  in this case the component fields do not transform canonically
under the deformed supersymmtery transformation.
In the singlet case, we will show that there is a field redefinition
such that the redefined fields also transform canonically under the deformed 
supersymmetry.
We will construct a deformed Lagrangian which is invariant 
under both the gauge and supersymmetry transformations.
We find that the deformed Lagrangian is characterized by a single function
of an anti-holomorphic scalar field.

%
%
We begin with reviewing the non(anti)commutative 
deformation of ${\cal N}=2$ harmonic
superspace \cite{IvLeZu,FeSo,ArItOh2}.
The ${\cal N}=2$ harmonic superspace \cite{GaIvOgSo} has coordinates
$(x^{\mu},\theta^{\alpha}_i, \bar{\theta}^{\dot{\alpha}i},u^{\pm i})$,
where $\mu=0,1,2,3$ are spacetime indices, $\alpha,\dot{\alpha}=1,2$
spinor indices and $i=1,2$ $SU(2)_{R}$ indices.
We will consider the Euclidean signature of spacetime.
For lowering and raising spinor indices, we use
an antisymmetric tensor $\varepsilon_{\alpha\beta}$
with $\varepsilon^{12}=-\varepsilon_{12}=1$, while
for $SU(2)_{R}$ indices, we use
$\epsilon_{ij}$ with $\epsilon^{12}=-\epsilon_{12}=-1$.
The coordinates $(x^{\mu},\theta^{\alpha}_i,
\bar{\theta}^{\dot{\alpha}i})$
are those of ${\cal N}=2$ rigid superspace.
The bosonic variables 
$u^{\pm i}$, called the harmonic variables, form an $SU(2)$ matrix satisfying
$ u^{+i}u^{-}_{i}=1$ and 
$\overline{u^{+i}}=u^{-}_{i}$.
The harmonic variables are necessary for the off-shell formulation of
supersymmetric field theories with extended supersymmetry as developed
in \cite{GaIvOgSo}.
The supersymmetry generators $Q^{i}_{\alpha}$,
$\bar{Q}_{\dot{\alpha}i}$ and the supercovariant
derivatives $D_{\alpha}^{i}$, $\bar{D}_{\dot{\alpha}i}$ are
defined by
\begin{eqnarray}
Q_{\alpha}^{i}&=&{\partial\over\partial\theta^{\alpha}_{i}}
-i(\sigma^{\mu})_{\alpha\dot{\alpha}}\bar{\theta}^{\dot{\alpha} i}
{\partial\over\partial x^{\mu}}, \quad
\bar{Q}_{\dot{\alpha} i}= -{\partial\over\partial\bar{\theta}^{\dot{\alpha} i}}
+i\theta^{\alpha}_{i}(\sigma^{\mu})_{\alpha\dot{\alpha}}
{\partial\over \partial x^{\mu}}
\nonumber\\
D_{\alpha}^{i}&=&{\partial\over\partial\theta^{\alpha}_{i}}
+i(\sigma^{\mu})_{\alpha\dot{\alpha}}\bar{\theta}^{\dot{\alpha} i}
{\partial\over\partial x^{\mu}}, \quad
\bar{D}_{\dot{\alpha} i}= -{\partial\over\partial\bar{\theta}^{\dot{\alpha} i}}
-i\theta^{\alpha}_{i}(\sigma^{\mu})_{\alpha\dot{\alpha}}
{\partial\over \partial x^{\mu}}.
\end{eqnarray}
In the harmonic superspace, we use the supercovariant derivatives
\begin{equation}
 D^{\pm}_{\alpha}=u^{\pm}_{i}D^{i}_{\alpha}, \quad
\bar{D}^{\pm}_{\alpha}=u^{\pm}_{i}\bar{D}^{i}_{\alpha},
\end{equation}
which are $U(1)$-projected by using $u^{\pm}_i$.
For the off-shell formulation of field theories,
the basic ingredient is the analytic superfield $\Phi$ satisfying
$ D^{+}_{\alpha}\Phi=\bar{D}^{+}_{\dot{\alpha}}\Phi=0$.
The solution of these constraints
 can be conveniently written of the form 
$\Phi=\Phi(x^{\mu}_{A},\theta^{+},\bar{\theta}^{+},u)$ by introducing 
analytic coordinates
\begin{eqnarray}
 x_{A}^{\mu}&=& 
x^{\mu}-i (\theta^i \sigma^\mu \bar{\theta}^j +\theta^j
\sigma^\mu\bar{\theta}^i)
u^{+}_{i}u^{-}_{j}
=x^{\mu}-i (\theta^{+}\sigma^{\mu}\bar{\theta}^{-}+\theta^{-}
\sigma^{\mu}\bar{\theta}^{+}),
\\
\theta^{\pm}_{\alpha}&=& u^{\pm }_{i}\theta^{i}_{\alpha},\quad
\bar{\theta}^{\pm}_{\dot{\alpha}}= u^{\pm}_{i}\bar{\theta}^{i}_{\dot{\alpha}}.
\end{eqnarray}

We now introduce the nonanticommutativity in the ${\cal N}=2$ harmonic
superspace by using the $*$-product:
\begin{equation}
 \{ \theta^{\alpha}_{i}, \theta^{\beta}_{j}\}_{*}=C^{\alpha\beta}_{ij},
\label{eq:nac1}
\end{equation}
with some constants $C^{\alpha\beta}_{ij}$.
We assume that the chiral coordinates 
$x_{L}^{\mu}\equiv x^\mu+i\theta_i \sigma^\mu \bar{\theta}^i$ 
and $\bar{\theta}_{\dot{\alpha}i}$
(anti-)commute with other coordinates
\begin{equation}
 [x^{\mu}_{L}, x^{\nu}_{L}]_{*}=[x^{\mu}_{L}, \theta^{\alpha}_{i}]_{*}=
[x^{\mu}_{L},\bar{\theta}^{\dot{\alpha} i}]_{*}=0, \quad 
\{ \bar{\theta}^{\dot{\alpha} i}, \bar{\theta}^{\dot{\beta} j}\}_{*}=
\{ \bar{\theta}^{\dot{\alpha} i}, \theta^{\alpha}_{j}\}_{*}=0.
\label{eq:nac2}
\end{equation}
Here the $*$-product realizing this non(anti)commutativity is defined by
\begin{equation}
 f*g(\theta)=f(\theta)\exp(P) g(\theta),\quad
P=-{1\over2}
\overleftarrow{Q^{i}_{\alpha}}
C^{\alpha\beta}_{ij}\overrightarrow{Q^{j}_{\beta}}.
\label{eq:moyal2}
\end{equation}
The Poisson structure $P$ commutes with the supercovariant derivatives. 
This deformation preserves chirality. 
The constants $C^{\alpha\beta}_{ij}$ is a symmetric
property $C^{\alpha\beta}_{ij}=C^{\beta\alpha}_{ji}$ and
can be decomposed of the form:
\begin{equation}
C^{\alpha\beta}_{ij}=C^{\alpha\beta}_{(ij)}
+{1\over4}\epsilon_{ij}\varepsilon^{\alpha\beta}C_{s}.
\end{equation}
Here the first term $C^{\alpha\beta}_{(ij)}$ is symmetric with respect 
to $i$ and $j$ (and also $\alpha$ and $\beta$).
The second term is antisymmetric and is called the singlet deformation
introduced in \cite{IvLeZu,FeSo}, 
In \cite{FeSo} the deformation with the Poisson structure
$P=-{1\over8}\varepsilon^{\alpha\beta}\epsilon_{ij}C_s
\overleftarrow{D^{i}_{\alpha}}\overrightarrow{D^{j}_{\beta}}$
has been studied.
In this paper we will consider the $*$-product (\ref{eq:moyal2})
with the singlet deformation parameter\cite{IvLeZu}: 
\begin{equation}
P=-{1\over8}\varepsilon^{\alpha\beta}\epsilon_{ij}C_s
\overleftarrow{Q^{i}_{\alpha}}\overrightarrow{Q^{j}_{\beta}}
. 
\end{equation}

The action of ${\cal N}=2$ supersymmetric $U(1)$ gauge 
theory in this non(anti)commutative harmonic superspace is written 
in terms of an analytic superfield $V^{++}$ \cite{Zu}:
\begin{equation}
 S={1\over2}\sum_{n=2}^{\infty}{(-i)^n\over n}
\int d^4 x d^{8}\theta du_{1}\cdots du_{n}
{V^{++}(\zeta_{1},u_1)*\cdots *V^{++}(\zeta_{n},u_n)
\over (u^{+}_{1}u^{+}_{2})\cdots (u^{+}_{n}u^{+}_{1})}
\label{eq:action1}
\end{equation}
where $\zeta_{i}=(x_{A},\theta^{+}_{i},\bar{\theta}^{+}_{i})$ and 
$d^8\theta=d^4 \theta^+ d^4\theta^- $ with
$d^4\theta^{\pm}=d^2\theta^{\pm}d^2\bar{\theta}^{\pm}$.
The harmonic integral $\int du$ is defined as in \cite{GaIvOgSo}. 
The action (\ref{eq:action1}) is invariant under the gauge transformation
\begin{equation}
 \delta^{*}_{\Lambda} V^{++}=-D^{++}\Lambda+i [ \Lambda, V^{++}]_{*},
\label{eq:gauge2}
\end{equation}
where the gauge parameter $\Lambda(\zeta,u)$ is also analytic.
$D^{++}$ denotes the harmonic derivative
$D^{++}
= 
	u^{+ i} {\partial \over \partial u^{- i} }
	- 2 i \theta^{+}\sigma^\mu \bar{\theta}^{+} 
		{\partial \over \partial x_A^\mu }
	+ \theta^{+ \alpha} {\partial \over \partial \theta^{- \alpha}}
	+ \bar{\theta}^{+ \dot{\alpha}} 
		{\partial \over \partial \bar{\theta}^{- \dot{\alpha}}}
$.
When an analytic superfield is expanded in the Grassmann coordinates,
each component field has a harmonic expansion with respect to
$u^{\pm}_i$.
The analytic superfield $V^{++}$ therefore contains
infinitely many auxiliary fields. 
Since the gauge parameter $\Lambda$ also includes infinitely many
fields, 
one can remove unnecessary auxiliary fields as in the commutative case.
We then arrive at the Wess-Zumino(WZ) gauge:
\begin{eqnarray}
 V^{++}_{WZ}(\zeta,u)
&=& 
-i\sqrt{2}(\theta^{+})^2 \bar{\phi}(x_{A})
+i\sqrt{2}(\bar{\theta}^{+})^2 \phi(x_{A})
-2i \theta^{+}\sigma^{\mu}\bar{\theta}^{+}A_{\mu}(x_{A})\nonumber\\
&&+4(\bar{\theta}^{+})^2\theta^{+}\psi^{i}(x_{A}) u^{-}_{i}
-4(\theta^{+})^2\bar{\theta}^{+}\bar{\psi}^{i}(x_{A})u^{-}_{i}
+3(\theta^{+})^2(\bar{\theta}^{+})^2 D^{ij}(x_{A})u^{-}_{i} u^{-}_{j}. 
\nonumber\\ 
\label{eq:wz1}
\end{eqnarray}
In \cite{ArItOh2}, we have computed the $O(C)$ action
(\ref{eq:action1}) in the WZ gauge explicitly.
In the singlet case, 
the order $O(C_s)$ Lagrangian reads 
${\cal L}= {\cal L}^{(0)}+{\cal L}^{(1)}$, where
\begin{eqnarray}
 {\cal L}^{(0)}&=&
-{1\over4}F_{\mu\nu}(F^{\mu\nu}+\tilde{F}^{\mu\nu})-i\psi^i \sigma^\mu
\partial_{\mu}\bar{\psi}_i -\partial^{\mu}\phi\partial_{\mu}\bar{\phi}
+{1\over4}D_{ij}D^{ij},\nonumber\\
{\cal L}^{(1)}&=&
{1\over\sqrt{2}}C_s A_{\nu}\partial_{\mu}\bar{\phi}
(F^{\mu\nu}+\tilde{F}^{\mu\nu})
+{i\over\sqrt{2}} C_{s} 
\bar{\phi} (\psi^k \sigma^{\nu}\partial_{\nu}\bar{\psi}_k)
+{i\over\sqrt{2}} C_{s} 
(\psi^k \sigma^{\nu}\bar{\psi}_k)\partial_{\nu}\bar{\phi}
\nonumber\\
&&-{i\over2}C_s \varepsilon^{\alpha\beta}A_{\mu} (\sigma^{\mu}\bar{\psi}^k)_{\alpha}
(\sigma^{\nu}\partial_{\nu}\bar{\psi}_k)_{\beta}
+{i\over2}C_{s}\bar{\psi}^i \bar{\psi}^j D_{ij}
+{\sqrt{2}\over4}C_{s}A_{\mu}A^{\mu}\partial^2\bar{\phi}
-{\sqrt{2}\over4}C_{s}\bar{\phi}D^{ij}D_{ij}, \nonumber\\
\label{eq:orderclagrangian1}
\end{eqnarray}
where $\tilde{F}_{\mu\nu} \equiv {i \over 2} \varepsilon^{\mu\nu\rho\sigma} 
F_{\rho\sigma}$. 
Note that ${\cal L}^{(0)}$ is the undeformed Lagrangian.

In the commutative case, the gauge transformation with 
the gauge parameter $\lambda(x_A)$ preserves 
the WZ gauge. But for generic $C^{\alpha\beta}_{ij}$, $\lambda(x_A)$ 
does not. 
In \cite{ArItOh2}, we have constructed the gauge parameter 
$\Lambda(\zeta,u)$ which preserves the WZ gauge, 
which is an infinite power series in the deformation parameter $C$.
We will see now the deformed gauge transformation in the singlet case more
explicitly.
For later convenience, we begin with the deformed gauge transformation 
(\ref{eq:gauge2}) of $V^{++}_{WZ}$ 
with the most general analytic gauge parameter $\Lambda(\zeta,
u) $: 
\begin{eqnarray}
\Lambda(\zeta, u) 
&=& 
	\lambda^{(0,0)} (x_A,u)
	+ \bar{\theta}^{+}_{\dot{\alpha}} \lambda^{(0,1)}{}^{\dot{\alpha}} (x_A,u)
	+ \theta^{+ \alpha} \lambda^{(1,0)}_{\alpha} (x_A,u)
	+ (\bar{\theta}^{+})^2 \lambda^{(0,2)} (x_A, u)
	\nonumber\\
&&{} 
	+ (\theta^{+})^2 \lambda^{(2,0)} (x_A, u)
	+ \theta^{+}\!\sigma^\mu \bar{\theta}^{+} 
		\lambda_\mu^{(1,1)} (x_A, u)
	+ (\bar{\theta}^{+})^2 \theta^{+}{}^{\alpha} 
		\lambda^{(1,2)}_{\alpha} (x_A, u)
	\nonumber\\
&&{} 
	+ (\theta^{+})^2 \bar{\theta}^{+}_{\dot{\alpha}} 
		\lambda^{(2,1)}{}^{\dot{\alpha}} (x_A, u)
	+ (\theta^{+})^2 (\bar{\theta}^{+})^2 
		\lambda^{(2,2)} (x_A, u) 
. 
\end{eqnarray} 
Here we have denoted the $(\theta^{+})^{n} (\bar{\theta}^{+})^{m}$-component 
as $\lambda^{(n,m)} (x_A, u)$. 
In the case of the singlet deformation, 
the gauge variation of $V^{++}_{WZ}$ corresponding to this general gauge parameter 
is calculated as 
\begin{eqnarray}
\delta^{*}_{\Lambda} V^{++}_{WZ} 
&=& 
	- \partial^{++} \lambda^{(0,0)} 
	+ \bar{\theta}^{+}_{\dot{\alpha}} 
		\left( 
		- \partial^{++} \lambda^{(0,1)}{}^{\dot{\alpha}} 
		\right)
	+ \theta^{+}{}^{\alpha} 
		\left( 
		- \partial^{++} \lambda^{(1,0)}_\alpha 
		\right) 
		\nonumber\\
&&{} 
	+ (\bar{\theta}^{+})^2 
		\left( 
		i C_s \partial_\mu \lambda^{(0,0)} A^\mu 
		- \partial^{++} \lambda^{(0,2)} 
		\right) 
	+ (\theta^{+})^2 
		\left( 
		- \partial^{++} \lambda^{(2,0)} 
		\right) 
		\nonumber\\
&&{} 
	+ \theta^{+} \sigma^\mu \bar{\theta}^{+} 
		\left( 
		2 i \partial_\mu \lambda^{(0,0)} 
		+ \sqrt{2} i C_s 
			\partial_\mu \lambda^{(0,0)} \bar{\phi}
		- \partial^{++} \lambda_\mu^{(1,1)} 
		\right) 
		\nonumber\\
&&{} 
	+ (\bar{\theta}^{+})^2 \theta^{+}{}^{\alpha} 
		\left( 
		- 2 C_s \partial_\mu \lambda^{(0,0)} 
			(\sigma^\mu \bar{\psi}^{i} )_\alpha u^{-}_{i}
		- i (\sigma^\nu \partial_\nu \lambda^{(0,1)})_\alpha 
		- {i\over \sqrt{2}} C_s 
			( \sigma^\nu \partial_\nu \lambda^{(0,1)} )_\alpha \bar{\phi} 
		\right.\nonumber\\
&&		\left.\quad\quad {} 
		+ {i \over 2} C_s \lambda^{(1,0)}{}^{\beta} 
			(\sigma^\mu 
			\bar{\sigma}^\nu \varepsilon)_{\beta\alpha} 
			\partial_\mu A_\nu 
		+ i C_s \partial_\mu \lambda^{(1,0)}_{\alpha} 
			A_\mu 
		- \partial^{++} \lambda^{(1,2)}_\alpha 
		\right) 
		\nonumber\\
&&{} 
	+ (\theta^{+})^2 \bar{\theta}^{+}_{\dot{\alpha}} 
		\left( 
		i \partial_\mu \lambda^{(1,0)}{}^\beta 
			\sigma^\mu_{\beta{\dot{\beta}}} 
			\varepsilon^{\dot{\beta}\dot{\alpha}} 
		+ {i \over \sqrt{2}} C_s \partial_\nu 
			\left\{ 
			\lambda^{(1,0)}{}^{\alpha}  
				\sigma^\nu_{\alpha{\dot{\beta}}} 
				\varepsilon^{{\dot{\beta}}{\dot{\alpha}}} 
				\bar{\phi}
			\right\} 
		- \partial^{++} \lambda^{(2,1)}{}^{\dot{\alpha}} 
		\right) 
		\nonumber\\
&&{} 
	+ (\theta^{+})^2 (\bar{\theta}^{+})^2 
		\left( 
		C_s \partial_\nu 
			\left\{ 
			\lambda^{(1,0)}{}^{\alpha} 
				(\sigma^\nu \bar{\psi}^{-})_{\alpha} 
			\right\} 
		- i \partial^\mu \lambda_\mu^{(1,1)} 
		\right.\nonumber\\
&&		\left. \qquad\qquad {} 
		- {i \over \sqrt{2}} C_s  
			\partial^\mu ( \lambda_\mu^{(1,1)} \bar{\phi} )
		+ i C_s \partial^\mu ( \lambda^{(2,0)} A_\mu )
		- \partial^{++} \lambda^{(2,2)}
		\right) 
, 
\label{eq:gengaugevar}
\end{eqnarray}
where
$
\partial^{++} 
\equiv 
	u^{+ i} {\partial \over \partial u^{- i} }
$. 
Requiring that the analytic gauge parameter preserves the WZ gauge, 
we find that the gauge parameter should satisfy 
$
\lambda^{(0,0)} (x_A, u) =  \lambda (x_A)
$ 
and the other components are zero.
Namely, the analytic gauge parameter retaining the WZ gauge 
is of the same form as in the commutative case: 
\begin{equation}
\Lambda(\zeta, u) = \lambda (x_A). 
\end{equation}
Then we immediately find 
the deformed gauge transformation laws for the component fields 
in the case of the singlet deformation: 
\begin{eqnarray}
&&
\delta^{*}_{\Lambda} A_\mu 
= 
- \Bigl( 1 + {1\over \sqrt{2}} C_s \bar{\phi} \Bigr) \partial_\mu \lambda 
, \quad 
\delta^{*}_{\Lambda} \phi 
= 
{1\over \sqrt{2}} C_s A_\mu \partial^\mu \lambda
, \quad 
\delta^{*}_{\Lambda} \psi^i_{\alpha}
= 
- \frac{1}{2} C_s \partial_\mu \lambda \ (\sigma^\mu \bar{\psi}^{i} )_\alpha
, \nonumber\\
&&
\delta^{*}_{\Lambda} \bar{\phi} 
= 
\delta^{*}_{\Lambda} \bar{\psi}^i_{\dot{\alpha}}
= 
\delta^{*}_{\Lambda} D^{ij} 
= 
	0
	. 
\label{eq:gaugetr}
\end{eqnarray}
Note that ther is no higher order correction in $C_s$. 

We will determine the supersymmetry transformation $\delta_{\xi}$ 
generated by the supersymmetry generators $Q^i_\alpha$. 
First we consider the action of $Q^i_\alpha$ 
on the gauge superfield: 
\begin{equation}
\tilde{\delta}_{\xi}  V^{++}_{WZ} 
\equiv \xi_i^{\alpha} Q^i_{\alpha} V^{++}_{WZ} 
. 
\end{equation}
In the analytic basis, 
\begin{equation}
\xi_i^{\alpha} Q^i_{\alpha} 
= 
	- \xi^{+\alpha}Q^{-}_{\alpha}
	+ \xi^{-\alpha}Q^{+}_{\alpha}
	, 
\end{equation}
where 
$
\xi^{\pm}_{\alpha} \equiv \xi^{i}_{\alpha} u^{\pm}_{i}
$
and 
\begin{equation}
Q^{+}_{\alpha}
=
	{\partial\over\partial\theta^{-\alpha}}
	-2i \sigma^{\mu}_{\alpha\dot{\alpha}}\bar{\theta}^{+\dot{\alpha}}
		{\partial\over\partial x^{\mu}_{A}}
		,
\quad 
Q^{-}_{\alpha}
= 
	- {\partial\over\partial \theta^{+\alpha}}
	. 
\end{equation} 
The variation is not affected by the nonanticommutativity, 
so that we have 
\begin{eqnarray} 
\tilde{\delta}_{\xi}V^{++}_{WZ}
&=& 
	\bar{\theta}^{+}_{\dot{\alpha}} 
		\left( 
		2 i (\xi^{+} \sigma^\mu)_{\dot{\beta}} 
			\varepsilon^{\dot{\beta}\dot{\alpha}} A_\mu 
		\right) 
	+ \theta^{+ \alpha} 
		\left( 
		- 2 \sqrt{2} i \xi^{+}_{\alpha} \bar{\phi} 
		\right) 
	+ (\bar{\theta}^{+})^2 
		\left( 
		4 \xi^{+} \psi^{i} u^{-}_{i} 
		\right) 
	\nonumber\\
&&{} 
	+ \theta^{+}\!\sigma^\mu \bar{\theta}^{+} 
		\left( 
		4 \xi^{+} \sigma_\mu \bar{\psi}^{i} u^{-}_{i} 
		\right) 
	+ (\bar{\theta}^{+})^2 \theta^{+}{}^{\alpha} 
		\left( 
		- 2 (\sigma^\mu \bar{\sigma}^\nu \xi^{-})_{\alpha} 
			\partial_\nu A_\mu 
		+ 6 \xi^{+}_{\alpha} D^{ij} u^{-}_{i}u^{-}_{j} 
		\right) 
	\nonumber\\
&&{} 
	+ (\theta^{+})^2 \bar{\theta}^{+}_{\dot{\alpha}} 
		\left( 
		2 \sqrt{2} (\xi^{-} \sigma^\mu)_{\dot{\beta}} 
			\varepsilon^{\dot{\beta}\dot{\alpha}} 
			\partial_\mu \bar{\phi} 
		\right) 
	+ (\theta^{+})^2 (\bar{\theta}^{+})^2 
		\left( 
		- 4 i \xi^{-} \sigma^\mu \partial_\mu \bar{\psi}^{i} u^{-}_{i} 
		\right) 
		. 
\end{eqnarray} 
This is out of the WZ gauge, 
so in order to retain the WZ gauge 
it should be associated with 
an appropriate deformed gauge transformation $\delta^{*}_{\Lambda}$, 
as in the commutative case. 
{}From the result for the deformed gauge variation with the most 
general gauge parameter, 
(\ref{eq:gengaugevar}), we can find that 
at least $\lambda^{(0,0)}$ and $\lambda^{(2,0)}$ should be zero. 
With the use of such an analytic gauge parameter, 
the equations to determine the deformed supersymmetric transformation 
of the component fields 
will be found from the equation 
\begin{equation}
\delta_{\xi}V^{++}_{WZ} 
= 
	\tilde{\delta}_\xi V^{++}_{WZ} 
	+ \delta^{*}_\Lambda V^{++}_{WZ} 
	. 
\end{equation}
{}From those equations, we can find the appropriate gauge parameter 
is the one as same as in the $C_s=0$ case
and determine the deformed supersymmetry transformations as 
\begin{eqnarray}
&& 
\delta_{\xi} A_\mu 
= 
	i \xi^{i} \sigma_\mu \bar{\psi}_{i}
, \quad 
\delta_{\xi} \phi 
= 
	- \sqrt{2} i \xi^{i} \psi_{i}
	, \quad 
\delta_{\xi} \bar{\phi}
= 
	0 
	, \nonumber\\
&& 
\delta_{\xi} \psi^{i}_{\alpha} 
= 
	\Bigl( 1 + {1\over \sqrt{2}} C_s \bar{\phi} \Bigr)   
		(\sigma^{\mu \nu} \xi^{i})_{\alpha} F_{\mu\nu}  		 
	- D^{ij} \xi_{\alpha j} 
	+ {1\over \sqrt{2}} C_s \xi^{i}_\alpha \partial_\mu \bar{\phi} A^\mu 
	, \nonumber\\
&& 
\delta_{\xi} \bar{\psi}^{\dot{\alpha} i}  
= 
	- \sqrt{2} (\bar{\sigma}^\mu \xi^{i} )^{\dot{\alpha}} 
		\Bigl( 1 + {1\over \sqrt{2}} C_s \bar{\phi} \Bigr)  \partial_\mu \bar{\phi} 
	, \nonumber\\
&& 
\delta_{\xi} D^{kl} 
= 
	- i \xi^{k} \sigma^\mu \partial_\mu 
		\left\{ 
		\bar{\psi}^{l} \Bigl( 1 + {1\over \sqrt{2}} C_s \bar{\phi} \Bigr)  
		\right\} 
	- i \xi^{l} \sigma^\mu \partial_\mu 
		\left\{ 
		\bar{\psi}^{k} \Bigl( 1 + {1\over \sqrt{2}} C_s \bar{\phi} \Bigr)  
		\right\} 
, 
\label{eq:susytr}
\end{eqnarray} 
where 
$\sigma^{\mu\nu} \equiv \frac14 
	( \sigma^\mu \bar{\sigma}^{\nu} - \sigma^\nu \bar{\sigma}^{\mu} )$. 

We have seen that the deformed gauge and supersymmetry transformations
are exact at the order $O(C_s)$.
One may consider the component action which is invariant under these
transformations.
The variation of ${\cal L}^{(1)}$ for the deformed gauge and
supersymmetry transformations produces new terms of order $O(C_s^2)$,
which should be cancelled by the variation of 
the $O(C_s^2)$ Lagrangian ${\cal L}^{(2)}$.
But it turns out that these deformed transformations change
only  the $\bar{\phi}$ dependence of interaction terms
among gauge fields and fermions. 
Since the spacetime coordinates have noncommutativity with nilpotent
parameters, we expect that 
the Lagrangian do not include the higher derivative terms.
Thus we assume the  Lagrangian takes the form
\begin{eqnarray}
 {\cal L}&=& f_{1}(\bar{\phi}) F_{\mu\nu} (F^{\mu\nu}+\tilde{F}^{\mu\nu})
+f_{2}(\bar{\phi}) \psi^i \sigma^{\mu}\partial_{\mu}\bar{\psi}_i
+f_{3}(\bar{\phi})\phi
+f_{4}(\bar{\phi}) D_{ij}D^{ij}
\nonumber\\
&&
+f_{5}(\bar{\phi}) A_{\mu} 
(\bar{\psi}^i \bar{\sigma}^{\mu}\sigma^{\nu}\partial_{\nu}\bar{\psi}_i)
+f_{6}(\bar{\phi}) A_{\mu} \partial_{\nu}\bar{\phi}
(F^{\mu\nu}+\tilde{F}^{\mu\nu})
+f_{7}(\bar{\phi}) A^{\mu}A_{\mu}
+f^{\mu\nu}_{8}(\bar{\phi}) A_{\mu}A_{\nu}
\nonumber\\
&& +f_{9}{}_{\mu}(\bar{\phi}) \psi^i \sigma^{\mu}\bar{\psi}_i
+f_{10\nu}(\bar{\phi})
 A_{\mu}\bar{\psi}^i \bar{\sigma}^{\mu}\sigma^{\nu}\bar{\psi}_i
+f_{11}(\bar{\phi}) D_{ij} \bar{\psi}^i \bar{\psi}^j
+f_{12}(\bar{\phi}) (\bar{\psi}^i \bar{\psi}^j)(\bar{\psi}_i \bar{\psi}_j).
\end{eqnarray}
Here $f_{i}$ are functions of $\bar{\phi}$ and its derivatives.
Invariance of the Lagrangian 
${\cal L}$ under the deformed gauge transformation (\ref{eq:gaugetr}) 
imposes the constrains on the functions $f_i$'s. 
Similarly invariance under the deformed supersymmetry transformation (\ref{eq:susytr}) 
leads to the further constraints on $f_i$'s. 
{}From the set of those constrains, 
$f_i$'s are solved by the function $f_2$.
Therefore the Lagrangian becomes
\begin{eqnarray}
 {\cal L}&=& i f_{2}(\bar{\phi})
\Bigl\{
-{1\over4} \Bigl(1+{1\over\sqrt{2}}C_s \bar{\phi}\Bigr)
F_{\mu\nu} (F^{\mu\nu}+\tilde{F}^{\mu\nu})
-i\psi^i \sigma^{\mu}\partial_{\mu}\bar{\psi}_i
+\Bigl(1+{1\over\sqrt{2}} C_s \bar{\phi}\Bigr)\partial^2\bar{\phi} \phi
\nonumber\\
&&
+{1\over4} {1\over 1+{1\over\sqrt{2}}C_s \bar{\phi}} D_{ij}D^{ij}
-{{iC_s \over2}
\over 1+{1\over\sqrt{2}}C_s \bar{\phi}}
A_{\mu} (\bar{\psi}^k \bar{\sigma}^{\mu}\sigma^{\nu} \partial_{\nu} \bar{\psi}_k)
-{1\over\sqrt{2}} C_s A_{\mu} \partial_{\nu} \phi
(F^{\mu\nu}+\tilde{F}^{\mu\nu})
\nonumber\\
&&
+{C_s\over 2\sqrt{2}}
\Bigl\{
\partial^{\mu}\partial_{\mu}\bar{\phi} -{{C_s\over \sqrt{2}}
\over 1+{1\over\sqrt{2}}C_s \bar{\phi}}\partial^{\mu}\bar{\phi}\partial_{\mu}\bar{\phi}
\Bigr\}A^{\mu}A_{\mu}
+{1\over4} {C_s^2 \over 1+{1\over\sqrt{2}}C_s \bar{\phi}}
\partial_{\mu}\bar{\phi} \partial_{\nu}\bar{\phi} A^{\mu}A^{\nu}
\nonumber\\
&&
+{i {C_s\over \sqrt{2}}\over 1+{1\over\sqrt{2}}C_s \bar{\phi}}
\partial^{\mu}\bar{\phi} \psi^k \sigma_{\mu}\bar{\psi}_k
+{i {C_s^2\over 2\sqrt{2}} 
\over \Bigl(1+{1\over\sqrt{2}}C_s \bar{\phi}\Bigr)^2}
\partial_{\nu} \bar{\phi} A_{\mu}\bar{\psi}^k \bar{\sigma}^{\mu}\sigma^{\nu}\bar{\psi}_k
\nonumber\\
&&
+{i\over2} {C_s \over \Bigl(1+{1\over\sqrt{2}}C_s \bar{\phi}\Bigr)^2}
D_{ij}\bar{\psi}^i\bar{\psi}^j
-{1\over4}
{C_s^2\over \Bigl(1+{1\over\sqrt{2}}C_s \bar{\phi}\Bigr)^3}
(\bar{\psi}^i\bar{\psi}^j)(\bar{\psi}_i \bar{\psi}_j)
\Bigr\}. 
\label{eq:LBeforeRedef} 
\end{eqnarray}
At the order $O(C_s)$, the Lagrangian reduced to the result
(\ref{eq:orderclagrangian1}) due to 
$f_2=-i\Bigl(1-{1\over\sqrt{2}}C_s \bar{\phi}\Bigr)+O(C_s^2)$.

Now we consider the field redefinition.
In the singlet deformation case, the $O(C_s)$ gauge transformation
(\ref{eq:gaugetr}) is exact.
We can redefine the component fields such that these transform
canonically under the deformed gauge transformation.
Let us introduce $\hat{A}_{\mu}$, $\hat{\phi}$ and $\hat{\psi}^i$ by
\begin{equation}
 \hat{A}_{\mu}= F(\bar{\phi})A_{\mu}, \quad 
\hat{\phi}= \phi+G(\bar{\phi})A_{\mu}A^{\mu}, \quad 
\hat{\psi}^i_{\alpha}= \psi^i
+H(\bar{\phi})A^{\mu} (\sigma^{\mu}\bar{\psi}^i)_{\alpha},
\label{eq:redef2}
\end{equation}
where
$F(\bar{\phi})$, $G(\bar{\phi})$ and $H(\bar{\phi})$ are functions of
$\bar{\phi}$.
If we require that these fields transform canonically, {\it i.e.}
$\delta_{\lambda}\hat{A}_{\mu}=-\partial_{\mu}\lambda$, 
and $\delta_{\lambda}\hat{\phi}=\delta_{\lambda}\hat{\psi}^i=0$,
then the functions $F$, $G$ and $H$ are determined as
\begin{equation}
 F(\bar{\phi})= {1\over 1+{1\over\sqrt{2}} C_s \bar{\phi}},
\quad 
 G(\bar{\phi})= {{1\over2\sqrt{2}} C_s
\over 1+{1\over\sqrt{2}} C_s \bar{\phi}},
\quad 
 H(\bar{\phi})= -{{1\over2} C_s\over 1+{1\over\sqrt{2}} C_s \bar{\phi}}.
\end{equation}
It is easy to see that 
the Lagrangian (\ref{eq:LBeforeRedef}) after the field redefinitions (\ref{eq:redef2}) 
also coincides with the $O(C_s)$ result in \cite{ArItOh2}. 

The ${\cal N}=2$ vector multiplet 
$(D^{ij},\hat{A}_{\mu},
\hat{\psi}^i,\bar{\psi}_i,\hat{\phi},\bar{\phi})$, however, 
does not transform canonically under the supersymmetry transformation.
But, if we instead perform field redefinitions as 
\begin{eqnarray}
&& 
a_\mu 
= 
	F(\bar{\phi}) A_\mu 
	, \quad 
\varphi 
= 
	F(\bar{\phi})^{2} 
	\left(
	\phi
	+ G(\bar{\phi}) A_\mu A^\mu 
	\right)
	, \quad 
\bar{\varphi}
= 
	\bar{\phi}  
	, \nonumber\\
&& 
\lambda^i_{\alpha}
= 
	F(\bar{\phi})^{2} 
	\left( 
	\psi^i_{\alpha} 
	+ H(\bar{\phi}) A_\mu 
		(\sigma^\mu \bar{\psi}^{i} )_\alpha
	\right) 
	, \quad 
\bar{\lambda}{}^{\dot{\alpha} i}
= 
	F(\bar{\phi})\bar{\psi}{}^{\dot{\alpha} i},
	\nonumber\\ 
&& 
\tilde{D}^{ij}
= 
	F(\bar{\phi})^{2} 
	\left(
	D^{ij}
	- 2 i H(\bar{\phi})\bar{\psi}^i\bar{\psi}^j
	\right)
	, 
\end{eqnarray}
we can show that the multiplet 
$(\tilde{D}^{ij},a_{\mu},
\lambda^i,\bar{\lambda}_i,\varphi,\bar{\varphi})$
now transforms canonically under the supersymmetry transformation 
as well as the gauge transformation: 
\begin{eqnarray}
&& 
\delta_{\xi}a_{\mu}= i\xi^i\sigma_{\mu}\bar{\lambda}_i, \quad 
\delta_{\xi}\varphi= -i\sqrt{2}\xi^i\lambda_i, \quad 
\delta_{\xi}\bar{\varphi}= 0 ,\nonumber\\
&& 
\delta_{\xi}\lambda^i_\alpha = 
(\sigma^{\mu\nu}\xi^i)_{\alpha} f_{\mu\nu}
-\tilde{D}^{ij}\xi_{j\alpha},
\quad 
\delta_{\xi}\bar{\lambda}{}^{\dot{\alpha} i}= 
 -\sqrt{2}(\bar{\sigma}^{\mu}\xi^i )^{\dot{\alpha}}
\partial_{\mu}\bar{\varphi}, \nonumber\\ 
&&
\delta_{\xi} \tilde{D}^{ij}= 
-i \left( \xi^i \sigma^{\mu}\partial_{\mu}\bar{\lambda}{}^j
+\xi^j\sigma^{\mu}\partial_{\mu}\bar{\lambda}{}^i
\right)
,
\end{eqnarray}
here $f_{\mu\nu} \equiv \partial_\mu a_\nu - \partial_\nu a_\mu$. 
In terms of these newly defined fields, the Lagrangian 
becomes
\begin{equation}
 {\cal L}
= 
i f_{2}(\bar{\varphi}) \Bigl(1+{1\over\sqrt{2}} C_s \bar{\varphi} \Bigr)^3
\Bigl\{
-{1\over4} 
f_{\mu\nu} (f^{\mu\nu}+\tilde{f}^{\mu\nu})
-i
\lambda^i \sigma^{\mu}\partial_{\mu}\bar{\lambda}_i
+ \varphi \partial^2\bar{\varphi}
+{1\over4}
\tilde{D}_{ij}\tilde{D}^{ij}
\Bigr\},
\end{equation}
which takes a simple form.

In this paper, we have determined the deformed gauge and supersymmetry
transformation of component fields of ${\cal N}=2$ supersymmetric $U(1)$
gauge theory in the noncommutative harmonic superspace with the singlet 
deformation parameter.
The Lagrangian which is invariant under these transformations are
obtained. 
In this work, we could not determine the complete Lagrangian 
due to the function $f_2(\bar{\phi})$, which is necessary for further study.
We have studied the field redefinition of component fields, such that these
fields transform canonically under the gauge transformation.
It is interesting to compare the present result to the Lagrangian in
\cite{FeSo}, which is based on the different Poisson structure.

It is also interesting to study the deformed supersymmetry in the case of 
non-singlet deformation. We expect that the action is invariant under
the deformed ${\cal N}=(1,0)$ supersymmetry.
In this case one may consider the ${\cal N}=1/2$ superspace limit by 
restricting
$C^{\alpha\beta}_{ij}$ to $C^{\alpha\beta}\delta^{1}_{i}\delta^{1}_{j}$.
The action in this limit is expected to
reduce to that in \cite{ArItOh1}, in which 
it was claimed that 
the ${\cal N}=2$ action has only ${\cal N}=1/2$ supersymmetry.
In \cite{ArItOh1}, the supersymmetry linearly deformed in $C$ 
has been examined. 
But the reduction from the harmonic superspace suggests that the
deformed supersymmetry is realized  nonlinearly in $C$.
In a subsequent paper \cite{ArItOh3}, we will study the structure
of supersymmetry for generic deformation parameters and clarify this
point.

{\bf Acknowledgments}:
We would like to thank A. Ohtsuka for useful discussion.
One of the authors (T. A.) is supported by 
the Grant-in-Aid for Scientific Research in Priority Areas (No.14046201) 
from the Ministry of Education, Culture, Science, Sports 
and Technology of Japan.

{\bf Note added}: 
After this paper was submitted to the e-archive, 
a new paper has appeared \cite{FeIvLeSoZu} 
where the undetermined function in the component Lagrangian 
is completely fixed.

\end{document}